# Single electron self-coherence and its wave/particle duality in the electron microscope

C. Kisielowski[1], P. Specht[2], J.R. Jinschek[3,4], S. Helveg[4]

1 The Molecular Foundry, Lawrence Berkeley National Laboratory, One Cyclotron Rd., Berkeley, CA 94720, USA
2 Department of Materials Science and Engineering, University of California Berkeley, Berkeley, CA 94720, USA
3 National Centre for Nano Fabrication and Characterization (DTU Nanolab), Technical University of Denmark, 2800 Kongens Lyngby, Denmark
4 Center for Visualizing Catalytic Processes (VISION), Department of Physics, Technical University of Denmark, 2800 Kongens Lyngby, Denmark
**\*** Correspondence: cfkisielowski@sbcglobal.net

**Abstract**

Intensities in high-resolution phase-contrast images from electron microscopes build up discretely in time by detecting single electrons. A wave description of pulse-like coherent-inelastic interaction of an electron with matter implies a time-dependent coexistence of coherent partial waves. Their superposition forms a wave package by phase decoherence of 0.5 - 1 radian with Heisenbergs energy uncertainty $\Delta E_H = \hbar/2 \, \Delta t^{-1}$ matching the energy loss $\Delta E$ of a coherent-inelastic interaction and setting the interaction time $\Delta t$. In these circumstances, the product of Planck's constant and the speed of light hc is given by the product of the expression for temporal coherence $\lambda^2/\Delta\lambda$ and the energy loss $\Delta E$. Experimentally, the self-coherence length was measured by detecting the energy dependent localization of scattered, plane matter waves in surface proximity exploiting the Goos-Hänchen shift. Chromatic-aberration Cc-corrected electron microscopy on boron nitride (BN) proves that the coherent crystal illumination and phase contrast are lost if the self-coherence length shrinks below the size of the crystal unit cell at $\Delta E > 200$ eV. In perspective, the interaction time of any matter wave compares with the lifetime of a virtual particle of any elemental interaction, suggesting the present concept of coherent-inelastic interactions of matter waves might be generalizable.

**Introduction**

The year 2024 marks the 100th anniversary of L. de Broglie's dissertation (de Broglie, 1924) introducing the duality of particles and matter waves by the wavelength $\lambda = h / mv$ where h is Planck's constant, m is the particle mass and v its velocity. Successively, this discovery stimulated the first manufacturing of an electron microscope (Ruska & Knoll, 1931). Ever since, instrument performance steadily improved to reach nowadays a spatial resolution beyond half the diameter of a hydrogen atom (Kisielowski et al., 2008) in high resolution transmission electron microscopy (HRTEM).

For decades, Dirk Van Dyck helped guiding the development and applications of the tool by establishing suitable theories to understand the contrast formation in the recorded HRTEM images (Van Dyck 1985, 2022). His view is consistent with the Copenhagen Convention of Quantum Mechanics (Bohr, 1958; Van Dyck et al., 2000) and formed the groundwork for rich progress.

Specifically, work at the National Center for Electron Microscopy (NCEM) in Berkeley, USA, benefitted from procedures to reconstruct electron exit wave functions (Coene et al. 1992, Kisielowski et al. 2001), the development of a S-state model for electron channeling (Geunes & Van Dyck, 2002), and the description of electron scattering as being mainly inelastic (Van Dyck et al. 2015). This contributed to a further development of discrete tomography (Jinschek et al., 2008) and the 'big bang' theory (Van Dyck et al., 2012) providing a fully quantitative description of atomic structures in 3D including electron beam-induced atom dynamics (Chen et al., 2016, 2021).

Today there are many opportunities for advancements, particularly in the context of quantum electron microscopy (Kruit et al., 2016) or electron microscopy with ultra-fast time resolution (Flannigan & Zewail, 2012; Feist et al., 2015, Kisielowski et al., 2019). Surprisingly, they are intertwined with Young's double slit experiment (Young, 1802), which laid the foundation for understanding the wave/particle duality of electrons and is often seen as a window into quantum physics. Traditionally, the observed intensity distribution of a double slit is explained by applying Huygens Principle (1690) to infinite, optical wave functions. It considers every point of a scattered wavefront to be the source of a secondary spherical wave and their envelop function describes the new wavefront (eg. Alonso M & Finn ER, 1983). The principle is also a basic building block for the successful deployment of multi-slice calculations to solve Schrödinger's static equation for matter waves (Cowley & Moody, 1957) because crystalline materials can be modeled by periodic arrays of empty channels and filled atom columns (Geunes & Van Dyck, 2002). Such calculations are widely applied to describe the coherent-elastic scattering of matter waves (e.g. Kilaas, 2024). However, the electron detection process is particle-like because an interference pattern is formed by detecting discrete electron scattering events over time (e.g. Tonomura et al., 1989) and it includes unsolved reservations about the concept of a wave function collapse. In particular, the validity of this wave/particle duality remains debated because the discrete nature of the intensity build-up from a double slit also occurs if photons are detected instead of electrons (Rueckner & Titcomb, 1996) and the intensity distribution can be described only by wave functions (Hobson, 2013) or only by particles (Parra, 2018) without the need for wave/particle duality.

Fortunately, technologies have evolved to the point where electron emissions can be manipulated one electron at a time and registered one by one to study single electron scattering events in electron microscopes with spherical and chromatic aberration correction (Haider et al., 2010). Hence, the build-up of image intensity by wave interferences in HRTEM images or diffraction patterns can be routinely studied in ultra-low dose rate conditions reaching 0.01 eÅ$^{-2}$s$^{-1}$ (Kisielowski et al. 2021). Statistically, only one electron is present in the microscope column during such experiments. Therefore, it is now feasible to investigate any discrete sequence of electron scattering events while striving to describe the resulting interference pattern in a wave description, as schematized in Figure 1. It distinctly complements the current state-of-the-art (Van Dyck et al., 2000) by considering, both, electron self-interferences and electron ensemble-interferences while it embraces a concept of decoherence to form wave packages with the width $l_s$ from infinite incident plane waves during time-dependent, coherent-inelastic interactions with the sample instead of invoking a wave function collapse at the detector.

Here, we discuss how the self-coherence length $l_s$ can be accessed experimentally and that it shrinks dramatically with increasing energy loss $\Delta E$ (Table I) while wavelength changes $\Delta\lambda$ are negligible. For Coulomb scattering, numerical $l_s$ values are large enough to coherently illuminate micro-fabricated double slits (Tavabi et al., 2017) or crystal structures to investigate the dependence of self-interferences on electron energy losses in energy-filtered EF-HRTEM experiments. Using the wave model of Figure 1 for any energy exchange $\Delta E = \hbar/2 \, \Delta t^{-1}$ in the Heisenberg limit during coherent-inelastic scattering processes, we derive the equation:

$$hc = (\lambda^2/\Delta\lambda) \, \Delta E \tag{1}$$

where h is Planck's constant and c is the speed of light. *Equation (1) reveals that the universal constant* hc *is given by the product of a temporal coherence length $\lambda^2/\Delta\lambda$ and the energy exchange $\Delta E$ of an inelastic interaction in the Heisenberg limit*. In this paper we show how the validity of this relationship can be tested using advanced phase contrast, chromatic-aberration Cc-corrected electron microscopy.

**Background**

Heisenberg's Uncertainty Principle is the central element in our considerations. The two forms

$$\Delta x \, \Delta p \geq \hbar/2 \tag{2}$$

$$\Delta E_H \Delta t \geq \hbar/2 \tag{3}$$

are commonly applied where $\hbar = h/2\pi$ is the reduced Planck constant, $\Delta t$ and $\Delta E_H$ are Heisenberg's time and energy uncertainties. $\Delta x$ and $\Delta p$ the corresponding position and momentum uncertainties, respectively. In electron optics, it is established practice to attribute the uncertainties of energy and position of the electron ensemble to the finite geometrical size $\Delta x_{gun}$ and the energy spread $\Delta E_{gun}$ of the electron gun. Their influence on image formation is modeled by integrating two damping functions A1 and A2 (aperture functions) to the contrast transfer function (CTF) of the microscopes, which describes the impact of spatial and temporal ensemble decoherence on resolution:

$$CTF(\mathbf{g}) = \sin(\chi(\mathbf{g})) * A_1 * A_2 \tag{4}$$

with $\chi(\mathbf{g}) = 1/2\pi \, (2f \, \lambda^2 \mathbf{g}^3 + C_s \lambda^3 \mathbf{g}^4 + O(\lambda)\ldots) \tag{5}$

$\mathbf{g}$ are spatial frequency vectors, Cs is the spherical aberration coefficient, f is the defocus and $O(\lambda)$ are higher order contributions. Thus, it is established knowledge that the coherence of the entire electron ensemble used for imaging (ensemble-coherence) limits the spatial resolution of HRTEM and considerable efforts are made to improve on the coherence of electron optics (Hawkes PW, 2023).

However, there is no exclusivity to set $\Delta E_H = \Delta E_{gun}$ in equation (3) because electron optical considerations for electron ensembles are not the only source of uncertainty. For example, in high energy physics interactions are explained in particle descriptions by exchanging a "virtual particle" of energy uncertainty $\Delta E$ and lifetime $\Delta t = \hbar/2 \ \Delta E^{-1}$ in the Heisenberg limit (e.g. Workman et al., 2022). Here, Figure 1 introduces a wave description for coherent-inelastic scattering by adapting any energy loss $\Delta E$ to the Heisenberg uncertainty $\Delta E_H = \Delta E = E_0-E_1$ and postulating the coexistence of coherent partial waves of wavelength spread $\Delta \lambda = \lambda_0(E_0) - \lambda_1(E_1)$ during the brief interaction time $\Delta t$. A restoration of the total electron wave function requires the superposition of n scattered partial wave functions (e.g. n=2 in Figure 1), which creates a wave package in real space (or a pulse in the time domain) because of the existing energy differences $\Delta E$ that cause related wavelength differences $\Delta \lambda$. Describing the loss of the phase relation among the partial waves by a decoherence phase $\varphi = 0.5$ radian gives:

$$0.5 = (2\pi/\lambda_0 - 2\pi/(\lambda_0+\Delta\lambda)) \ l_s \qquad (5)$$

In Figure 2, the calculated $l_s$ describes the energy dependent width of the created wave packages for $\varphi = 0.5$ radian (black line in figure 2), which is linked to the Heisenberg equation (3) by dividing $l_s$ with the speed of light c to obtain the self-coherence time $t_s = l_s/c$ (red line in Figure 2, Kisielowski et al., 2021). The self-coherence time $t_s$ can be interpreted as the interaction time $\Delta t$. Substituting $l_s = \hbar c / 2\Delta E$ in equation (5) yields $hc \approx (\lambda^2/\Delta\lambda) \ \Delta E$ omitting the term $(1+\Delta\lambda/\lambda)$. If it attributed to the phase decoherence $\varphi = \varphi(\lambda)$, $\varphi(\lambda)$ varies between 0.5 and 1 radian and the equality in equation (1) is obtained, which defines self-coherence.

Experimentally, values for the self-coherence length $l_s$ (blue squares in Figure 2) were determined by EF-HRTEM measurements with the chromatic-aberration Cc-corrected TEAM I microscope detecting the exponential decay of an evanescent field at an abrupt sample/vacuum interface (Kisielowski et al. 2023). The measured decay lengths show the close agreement of calculated and measured self-coherence length in Figure 2. It is critical for the experiments that a parallel electron beam of low convergence angle ($\alpha = 0.09$ mrad) is used because this allows maintaining a grazing incident angle in conditions of total reflection, where the coherently illuminated sample area determines the extension of the measured transversal evanescent field caused by the Goos-Hänchen shift (Carter & Hora, 1971, Kisielowski et al. 2023). In addition the presence of a Cc-corrector is needed to maintain a high spatial resolution in EF-HRTEM because it eliminates a focus dependence from the energy loss region ΔE < 600 eV (Haider et al., 2010). As previously reported (Kisielowski et al., 2023), slight differences between measurement and calculation in Figure 2 can be explained by the lower speed of the accelerated electrons at 300 kV of v = 0.77 c and a minor non-linearity of the average refraction index n=1.4 for GaN in the measured energy range. As for previously published literature data, it is satisfactory that our measurements exactly reproduce the available delocalization measurement by electron holography at 300 kV (Röder & Lichte, 2011) and systematically exceed the preceding delocalization measurements by scanning transmission electron microscopy (STEM) (Rez et al., 2017; Egerton 2017) because they were acquired by lower electron acceleration voltages (60 kV - 100 kV) with larger beam convergence angles around 20 mrad (gray symbols in Figure 2). A reduction of the acceleration voltage to 80 kV reduces the relativistic electron velocity to v = 0.50 c and the measured $l_s$ values by a factor v(80 kV)/v(300 kV) = 0.65, which is neglected in this work because it does not alter our conclusions.

**Measurements and results**

Here, we present the verification that ls determines the ability to form the interference pattern of HRTEM images. Our experiments used the Cc-corrected TEAM I microscope (Dahmen et al., 2009). Its electron optical parameters are given by Tiemeyer et al. and they provide a parallel electron beam of only 0.09 mrad beam convergence in a Nelsonian illumination scheme (Tiemeyer et al., 2012). The illumination takes advantage of the low energy spread $\Delta E_{gun} = 0.1$ eV of a monochromator and irradiates a field of view of only $10^5$ nm². Such electron optical parameters provide an achievable resolution of better than 0.5 Å for electron ensembles (Kisielowski et al., 2008).

We acquired focus series of 10 - 30 images in EF-HRTEM mode at 80 keV from free-standing hexagonal boron nitride (BN) along the c-axis (space group #186, a = 0.25 nm, c = 0.67 nm) Specific energy losses ΔE were selected from the energy loss range 0 eV ≤ ΔE ≤ 410 eV for recording the EF-HRTEM image series. To recover exit wave functions, equidistant defocus steps $f_s = -2.18 \pm 0.04$ nm generated the required phase relation between successive images of a recorded image series (equation 5) to apply the Gershberg & Saxton (1972) algorithm. All computation and analysis were performed by the Tempas software package (Kilaas, 2024).

The BN sample itself was a few monolayers thick < 7 nm, which ensures that multiple scattering is absent. Moreover, the choice of the specific energy losses ΔE marked by triangles in Figure 2 allows studying how phase contrast is affected when fine-tuning the electron self-coherence length $l_s$ to approach values close to the lattice parameter a = 0.25 nm of the BN crystal.

Figure 3 shows single EF-HRTEM images from the focus series and their Fourier Transform acquired with the zero-loss beam (a), at the low loss region (b), at the plasmon peak (c), at the boron K-edge (d), and the nitrogen K-edge (e), respectively. In case of imaging with the zero-loss beam (Figure 3a), the presence of unscattered electrons creates a bright field image while all other images (Figures 3b-e) are dark field images. A zero-loss beam width of 0.7 eV sets a resolution of ~ 1.2 Å at 80 kV for electron ensembles by equations (4) and (5) while the self-coherence lengths would vary between infinity and 100 nm (Figure 2). Hence, under such conditions ensemble-coherence and self-coherence coexist (degradation) with a resolution defined by an infinite plane wave. In contrast, EF-HRTEM at ΔE > 0 eV lifts this degeneracy.

When visually inspecting the images in Figures 3a to 3c, it is, however, challenging to recognize that different energy losses were used for image acquisition: all images exhibit the distinctive features of phase contrast imaging including the formation of Fresnel fringes or fringes from the strain fields around vacancies. The presence of these features in the loss region is known as an "extension of phase contrast into the loss region" (Kabius et al., 2009; Forbes et al., 2014). They become less pronounced in images from the boron K-edge at ΔE = 200 eV and cannot be recognized in the image taken from the nitrogen K-edge at ΔE = 410 eV where a long exposure time (10 s) is used while maintaining a high spatial resolution at a finite sample drift. The fading contrast in images from the core loss region already suggests that the self-coherence length drops below the size of the BN lattice parameter (Figure 2) where the coherent illumination of the crystal unit cell fails.

Simulating the image contrast in the images recorded at the nitrogen K-edge (N-core loss: ΔE = 410±12 eV, Figure 3e) is key to prove that coherence is lost and not impacted by noise limitations. The argument exploits the difference between linear **(0, g$_i$)** interferences of scattering vectors from an entire crystal unit cell and the detection of non-linear ($g_i$, $g_j$) cross transfer interferences of densely spaced lattice planes that have a shorter distance than the length of a unit cell vector (Bals et al., 2005). Relevant multi-slice simulations of thickness-defocus maps - still assuming coherent-elastic image formation - are shown in Figure 4. The linear interferences (**0, $g_i$**) dominate the varying intensity pattern of the simulated HRTEM images in Figure 4a that are used for exit wave reconstructions where diffraction spots are sampled to the resolution limit (Figure 4a'). However, to simulate the non-linear images of Figure 4b and compare with the experimental EF-HRTEM image of Figure 3e we need to insert an opaque aperture of 0.7 Å$^{-1}$ radius that matches the self-coherence length $l_s^{-1}$ = 0.7 Å$^{-1}$ at the energy loss of $\Delta E$ = 410 eV. The aperture excludes low spatial frequencies from diffraction (Figure 4b') and creates the characteristic image pattern of non-linear ($g_i$, $g_j$) interferences. The match of experiment and simulation in Figure 4c shows indeed that the self-coherence length $l_s$ acts as an aperture for low spatial frequencies even though they are detected in the Fourier transform of Figure 4c'. Hence, ensemble-coherence sets the resolution at high spatial frequencies while self-coherence modulates the low frequency limit at high energy losses.

In general, the match of experiment with simulation in Figure 4c implies the existence of a decoherence phase of 0.5 - 1 rad. Certainly, non-linear imaging occurs in any material (Bals et al., 2005). For example, multi-slice simulation of non-linear interferences explain the published EF-HRTEM images of core losses from strontium titanate ($SrTiO_3$) more convincingly than the introduction of a double channeling transition potential as previously published (Forbes et al., 2014).

Further support for our argument that the phase contrast depends on the self-coherence length is obtained by reconstructing the phase of the electron exit wave function from EF-HRTEM focal series that were recorded at the various energy losses (Figure 5). Figures 5a through 5d depict the reconstructed phase images of BN showing the atomic structure of BN with a dumbbell separation of 0.145 nm. This image resolution is maintained up to at least $\Delta E$ = 27 eV of energy loss but noise becomes relevant for a determination of resolution in images recorded at $\Delta E$ = 200 eV (boron K-edge). Exit wave reconstructions from images recorded at $\Delta E$ = 410 eV are not yet possible because at $\Delta E$ > 200 eV the value of the self-coherence length $l_s$ shrinks below the BN lattice parameter, therefore only non-linear ($g_i$, $g_j$) interferences are present and current reconstruction algorithms required the presence of linear interferences (**0**, **gi**). An extraction of line profiles (Figure 5d) reveals a monotonic decrease of the phase contrast difference $\Delta\Phi$ (max to min) with increasing energy loss that obeys a power law including a quadratic decay with inverse energy losses (Figure 5e). Such behavior is to be expected if the square of the self-coherence length $l_s^2$ defines the coherently illuminated sample area.

The result also supports our empirical finding that the use of a ~ $10^5$ nm² small area for sample irradiation in the Nelsonian illumination scheme with $\Delta E_{gun} = 0.1$ eV produces best results for phase contrast imaging (Kisielowski et.al, 2021). In this case, the square of the self-coherence length from Figure 2 is at least $l_s^2 = 6 \times 10^5$ nm² in size, which ensures that in our experiments with the TEAM I microscope the entire irradiated sample area is self-coherently illuminated.

**Relevance**

Surely, material sciences benefit from a better understanding of wave function localization in pulse-like energy exchanges during interactions with samples. For example, a structured temporal illumination with picosecond time resolution impacts beam-sample interactions in electron microscopy (Kisielowski et al., 2019) and can explain the retardation of beam-induced damage by time-temperature transformations. Moreover, self-coherence enables a critical dose estimate for the onset of beam-induced damage treating radiation soft and hard matter in exactly the same manner (Kisielowski et al., 2021).

Here, however, we focus on the relevance of our measurements to quantum mechanical aspects. Specifically, we address the wave/particle duality of matter waves, using electron waves as our model system. Principally, both particle and wave descriptions provide excellent explanations for physical processes in nature, while contradicting conclusions only arise when both pictures are mixed. Undoubtedly, the great success of field theories rests on the use of wave descriptions (Hobson, 2013) but there remain many unanswered questions. Hence, we seek a better understanding of the equivalence of wave and particle descriptions and what phenomena - if any - may only be explained by one of them.

In standard applications of electron microscopy, it often suffices to only consider the ensemble coherence of electron optics: For example, in multi-slice calculation an electron wave of 2 pm wavelength oscillates 10 000 times before exciting a 20 nm thick crystal and a decoherence phase of $\varphi = 0.5$ rad produces an equivalent wave package width of 20 nm at an energy loss of 5 eV (Figures 1, 2). Thus, inelastic scattering may be ignored for scattering events with $\Delta E < 5$ eV in HRTEM because wave packages cannot be distinguished from infinite wave functions. In fact, a scattered wave package of energy loss $< 10^{-7}$ eV would extend over a distance of about 1 meter and ensure entanglement of sample and detector in the same way as would be the case with coherent-elastic scattering ($\Delta E = 0$). Thus, it is established praxis to choose a coherent-elastic description of electron scattering in combination with a time-independent Schrödinger equation. The restricted validity of this reasonable approximation, however, becomes obvious in experiments that explore the time domain (Kisielowski et al., 2021, 2023) and if the degeneration of ensemble-coherence and self-coherence from the zero-beam region is lifted by recording in the energy loss region, as shown in this paper.

Our approach (Figure 1) addresses pulse-like Coulomb interactions in a wave picture. In literature, however, the Coulomb force is only one of the four fundamental forces. In fact, forces and interactions are commonly described in particle pictures, often in combination with Feynman diagrams where the exchange of a virtual particle mediates the interaction. They are bosons for weak interactions, gluons for strong interactions, virtual photons for coulomb interactions, and gravitons for gravitational interactions (Workman et al., 2022). A comparison with our results mandates expanding the scale of Figure 2 substantially. The scale expansion in Figure 6 is truly bold but intriguingly because the Heisenberg relation is a mathematical necessity that does not depend on the specific nature of any interaction and the decoherence phases $\varphi$ of 0.5 - 1 radian is set by the universal constant hc = 1.24 $10^{-6}$ eVm, which is valid for any coherent-inelastic interaction.

The apparent validity of equation (1) is revealed by comparison with existing literature (in Fig 6): By default, the Heisenberg limit is used for the determination of virtual particle lifetimes. For example, experimental data exist for the energy (80.4 GeV) and the energy uncertainty (2.1 GeV) of the W-boson (Workman et al., 2022) yielding $\Delta E/E \sim 3\%$. The Heisenberg limit links its energy uncertainty to its accepted lifetime of $10^{-25}$ s (Figure 6a). Data about gluons are hardly available but Wang & Boyanovsky (2001) reported the lifetime of a quark-gluon plasma of ~ 3 $10^{-23}$ s, which gives an energy uncertainty of 12.5 MeV that is 6% of the provided total energy value of 200 MeV (Figure 6a). An independent measurement of energy uncertainty and lifetime of a virtual photon was successful by its detection at 0.11 eV of energy, which allows estimating a ΔE/E ~ 5% energy uncertainty of 5 meV (de Aguiar Júnior et al., 2020) to give a lifetime of 50 fs. The time resolution of their instrument was limited to Δt > 400 fs. Nonetheless, it closely confirms the lifetime estimate of a virtual photon by the Heisenberg limit (Figure 6a). Remarkably, ultrafast measurements open up the possibility of independently measure energy uncertainty and lifetime, even in conditions that approach or even exceed the Heisenberg limit (Feist et al., 2015). To include gravitational interactions, the considered scale needs further expansion (as shown in Figure 6b) because the ratio of the coupling constants between e.g. gravitational and strong interactions is tiny ($10^{-39}$). Currently, the graviton is a theoretical construct of energy that is estimated to be smaller than $6 \cdot 10^{-32}$ eV (Olive et al., 2014) with an even smaller energy uncertainty. Therefore, when applying the Heisenberg limit, a lifetime of ≈ $10^{17}$ s can be expected.

Figure 6 highlights the equivalence of particle lifetime and the interaction time, which is interpreted in a wave picture as the duration of an interaction in the Heisenberg limit. It is astounding how many quantum-mechanical aspects are touched upon by establishing this equivalence: First, the gradual localization of wave functions with energy loss eliminates discussing their abrupt collapse. Both, the lifetime of virtual particles and the self-coherence of partial waves vary by more than 40 orders of magnitude because the Heisenberg limit does not depend on the nature of inelastic interactions. However, decoherence only exists in a wave picture and provides the expression for the temporal self-coherence length $l_s$ of equation (1). This remarkable length scale characterizes the range of possible interferences and entanglements and can be compared to distinctly different objects (Figure 6a): Interferences driven by weak interactions occur within the size of a proton; the strong interaction includes interferences of the size of atom nuclei; and the Coulomb interaction applies to the size of atoms and above, which is why it can be investigated by electron microscopy. Therefore, the localization of interferences by coherent-inelastic scattering is consistent with the established range of particle interactions and the small width of the wave packages at high energy loss allows them to be perceived as particles with wave/particle duality by humans. Understandably, the gravitational interaction is perceived to be fundamentally different because the self-coherence length vastly exceeds human dimensions in space and time (Figure 6b). Specifically, the interaction time in the Heisenberg limit is no longer a "brief" moment in time but reaches the age of the universe. Despite its tiny strength if compared to the other fundamental interactions, it must still produce self-coherences in a wave picture. They entangle the universe in space and time by forming wave packages that propagate at the speed of light because the Heisenberg relation remains valid. Gravitational waves have already been detected that originated somewhere in the universe (Weiss, 2018).

**Summary and Conclusion**

In addition to being an indispensable tool in material sciences, electron microscopy is uniquely suited to study fundamental quantum mechanical aspects including a pulse-like description of energy exchange in the Heisenberg limit $\Delta E = \hbar/2\, \Delta t^{-1}$ that forms wave packages by decoherence and yield equation (1). Self-coherence is the connection between the time-dependent Schrödingers equation (Kisielowski et al. 2023) and Heisenberg's energy uncertainty. The time uncertainty $\Delta t$ can be interpreted as an interaction time in a wave picture or a lifetime in a particle picture. This equivalence rationalizes the wave/particle duality. The width of wave packages is described by a self-coherence length $l_s$ that can vary by over 40 orders of magnitude. For Coulomb interactions, $l_s$ explains the vanishing phase contrast from EF-HRTEM images when the energy loss exceeds values of $\Delta E > 200$ eV. A procedure to experimentally determine the self-coherence length was established earlier by measuring the energy dependent localization of scattered matter waves in surface proximity. Therefore, electron microscopy can be used as a general tool to measure the self-coherence length in real space images.

Importantly, our interpretation is consistent with the established praxis that the formation of intensities in high-resolution images can be approximated by multi-simulations assuming coherent-elastic scattering of electron ensembles because in the case of zero-loss images ($\Delta E = 0$ eV) ensemble coherence and self-coherence coexist.

Technologically, the integration of a monochromator, chromatic aberration-corrector, and a direct electron detector into one instrument, provided by the TEAM project (Dahmen et al. 2009), made these measurements possible. Further emerging technologies, e.g. ultrafast electron microscopy, open opportunities to independently measure uncertainties in energy and in time. Concerning theoretical considerations, multi-slice calculations do not yet address decoherence or self coherence despite operating on wave functions.

**Acknowledgements**

The authors are thankful for many years of stimulating discussions and scientific explorations with Dirk van Dyck and congratulate him for his extraordinary career in the field of electron microscopy. Further, the authors acknowledges stimulating discussions with Dirk Van Dyck and Hannes Lichte in preparation of this paper. The Center for Visualizing Catalytic Processes is sponsored by the Danish National Research Foundation (DNRF146). Electron microscopy was funded by the Molecular Foundry, which is supported by the Office of Science, the Office of Basic Energy Sciences, the U.S. Department of Energy under Contract No. DE-AC02-05CH11231.

**Table I:**

The impact of energy loss on non-relativistic wavelength and self-coherence length for an electron acceleration by 300 kV.

| Voltage eV | Energy loss eV | Wavelength pm | Self-coherence length nm |
|---|---|---|---|
| 300000 | 0 | 2.239284 | infinity |
| 299999 | 1 | 2.239287 | 200 |
| 299990 | 10 | 2.239321 | 20 |
| 299900 | 100 | 2.239657 | 2 |

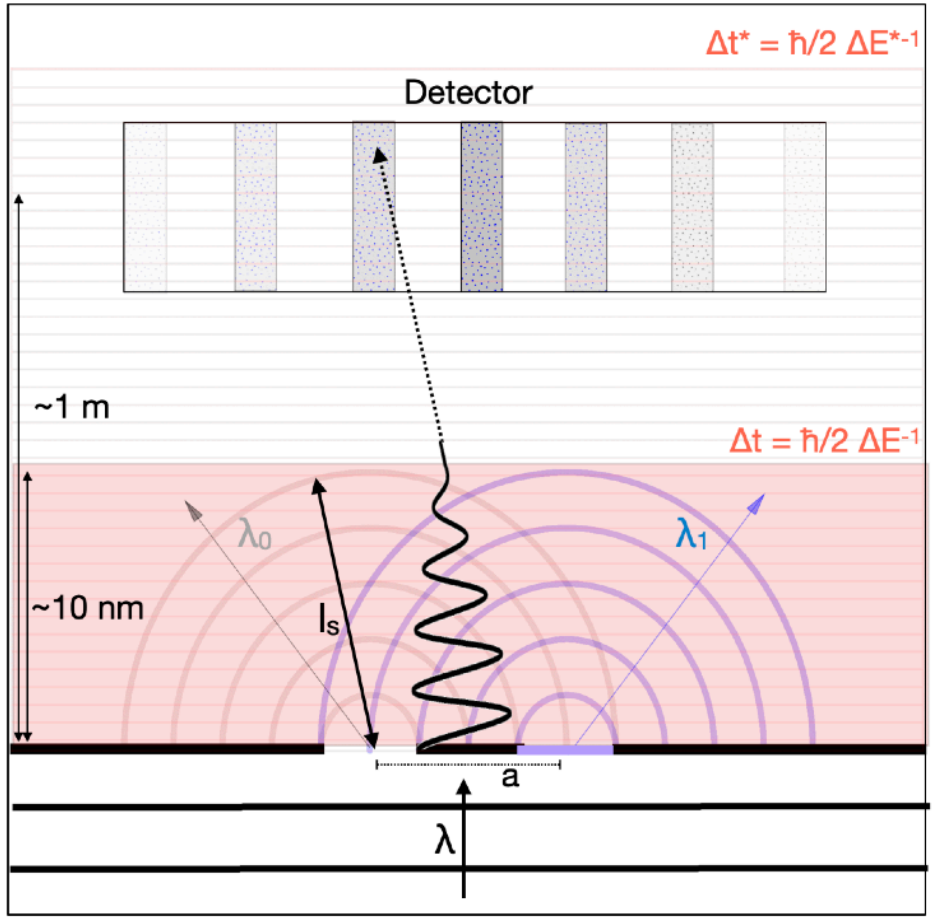

.Figure 1: Intensity builds up at a double slit by a time sequence of pulse-like, coherent-inelastic interactions of single matter waves.

An incoming matter wave of wavelength $\lambda$ (black) experiences an energy loss $\Delta E = E_0 - E_1$. The pulse-like interaction occurs at the object site in the Heisenberg limit $\Delta E = \hbar/2\ \Delta t^{-1}$. During the related interaction time $\Delta t$ (red space), Huygens Principle describes the resulting envelope function by the interference of the two coherent, secondary partial waves of wavelengths $\lambda_0(E_0)$, gray and $\lambda_1(E_1)$, blue. Their coherent superposition creates a wave package of width $l_s$ by decoherence (black), which is set by the number of secondary wave oscillations (5 are assumed). $l_s$ has longitudinal and transversal components. a is an atom column spacing or a lattice parameter. 10 nm is a typical sample thickness for HRTEM and 1 m is a typical sample-detector distance. Low loss Coulomb interactions yield $l_s$ values comparable to sample thickness during $\Delta t$. Sample and detector are entangled during the interaction time if $\Delta E^* \leq 10^{-7}$ eV as shown in this paper (red line texture).

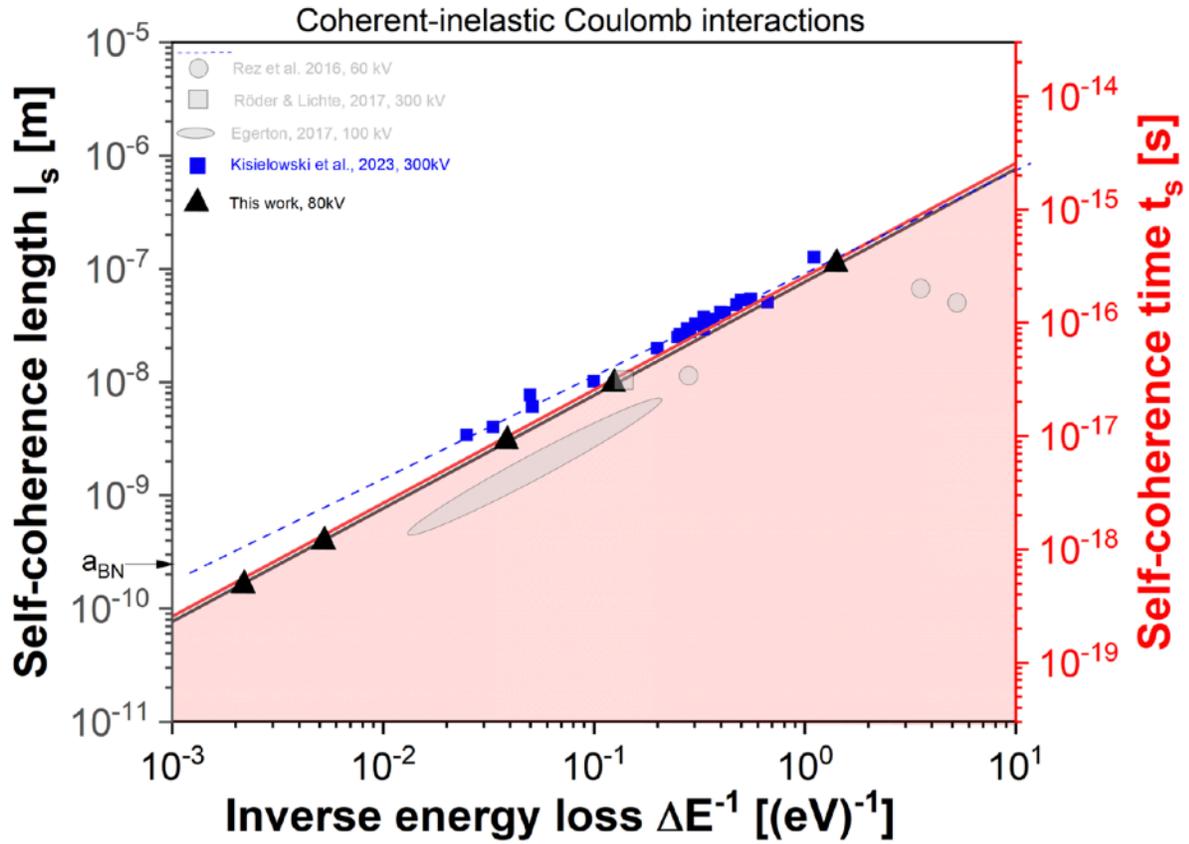

Figure 2: Energy dependence of the electron self-coherence length $l_s$ and self-coherence time $t_s = l_s/c$ in the Heisenberg limit for a decoherence phase of 0.5 radian (Kisielowski et al. 2021). The red area separates quantum space from real space (white). Literature data as indicated. The black triangles mark energy losses where focus series of EF-HRTEM images are acquired in this work. The a-lattice parameter of BN is indicated.

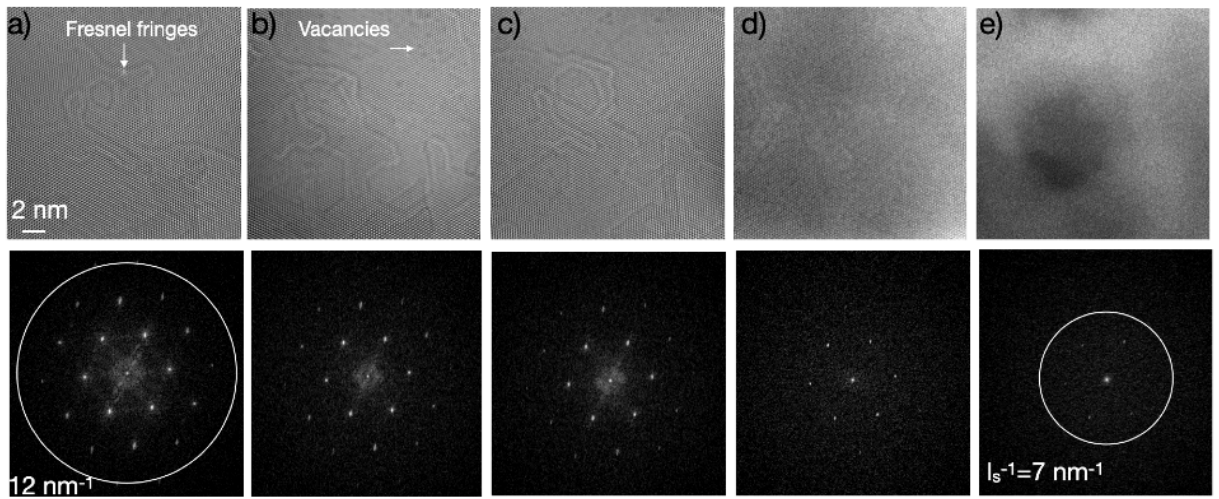

Figure 3: Single EF-HRTEM images and their Fourier Transform acquired at 80 kV across the energy loss region. (a) Zero loss, $\Delta E_{gun}$ = 0.7 eV, $\infty > l_s \geq 100$ nm, 645 e/pixel. The resolution limit is indicated (b) Low loss, $\Delta E$ = 8±4 eV, $l_s$ = 10 nm, 245 e/pixel, (c) Plasmon loss, $\Delta E$ = 27±4 eV, $l_s$ = 2.5 nm, 495 e/pixel, (d) B-core loss, $\Delta E$ = 200±12 eV, $l_s$ = 0.35 nm, 385 e/pixel, (e) N-core loss: $\Delta E$ = 410±12 eV, $l_s$ = 0.15 nm, 9 e/pixel. The inverse self-coherence length is indicated.

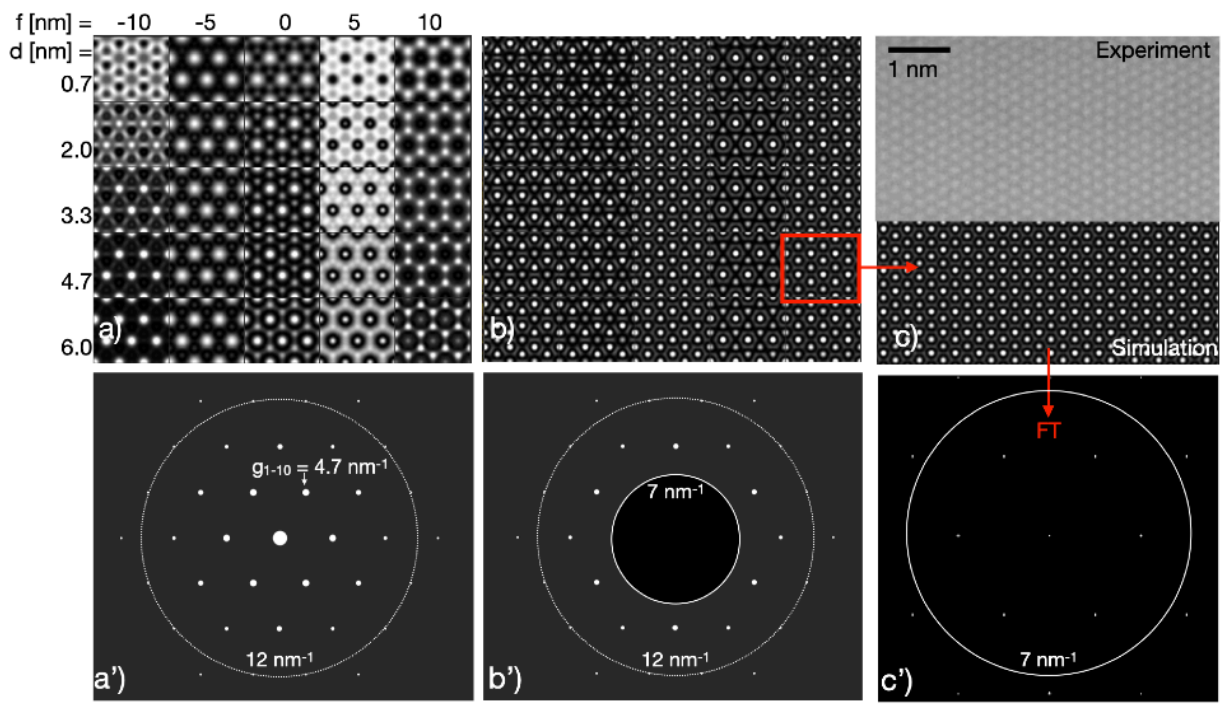

Figure 4: Multislice calculation of thickness (d) / defocus (f) maps of BN, 80 kV. a) Thickness/ defocus map of boron nitride and a') simulated diffraction pattern with the resolution of 12 nm$^{-1}$ indicated by dotted line. b) Non-linear image simulations (Bals et al., 2005) by inserting an opaque aperture of 1.5 Å (= 7 nm$^{-1}$) in the diffraction plane (full line in b'). c) The non-linear simulation is matched to the Fourier filtered EF-HRTEM image of Figure 3e recorded with $l_s$= 1.5 Å. The Fourier transform c' of the simulation c exhibits the low spatial frequencies that were excluded from detection by the opaque aperture in the diffraction plane of Figure 4b'.

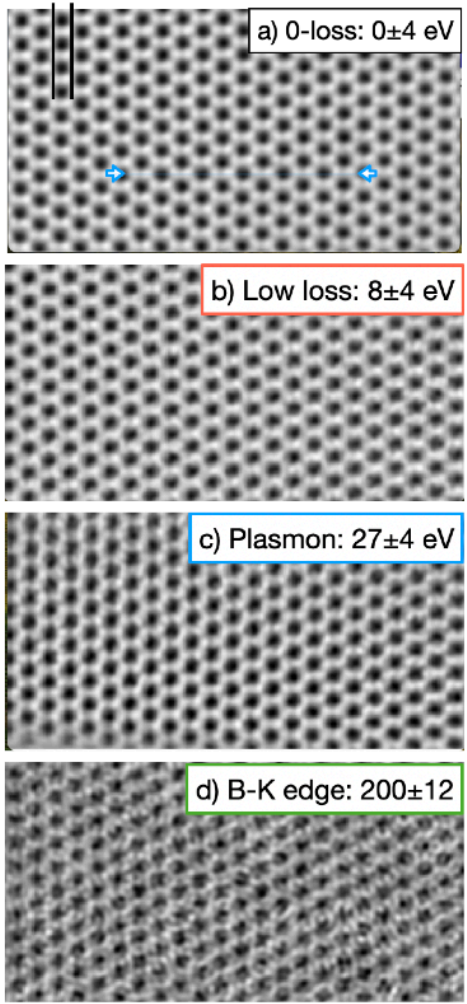
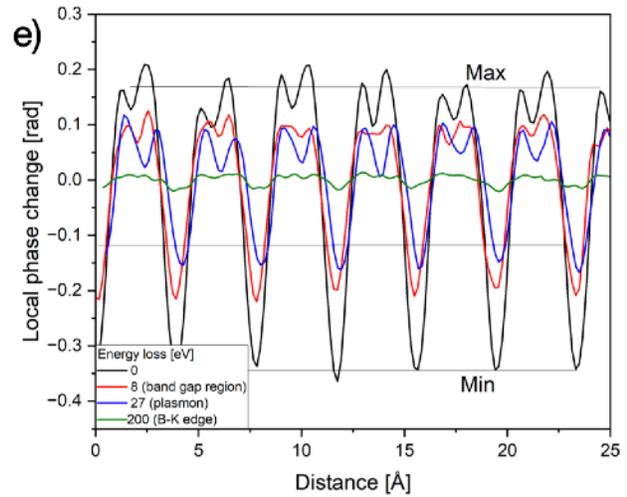
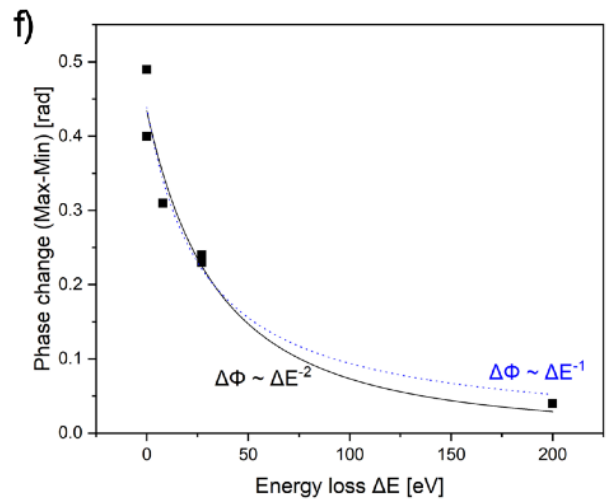

Figure 5: EF-HRTEM electron exit wave functions of BN, reconstructed from image focal series recorded at energy losses of a) 0 eV, b) 8 eV, c) 27 eV, and d) 200 eV (at the B-K edge), respectively. a-d) reconstructed phase images showing the atomic structure of BN with no evidence for resolution loss. e) Line profiles extracted from the measurements as indicated in Figure 5a. Local phase changes ΔΦ vary between an average minimum and a maximum phase value. f) The decay of the measured ΔΦ obeys a power law as indicated.

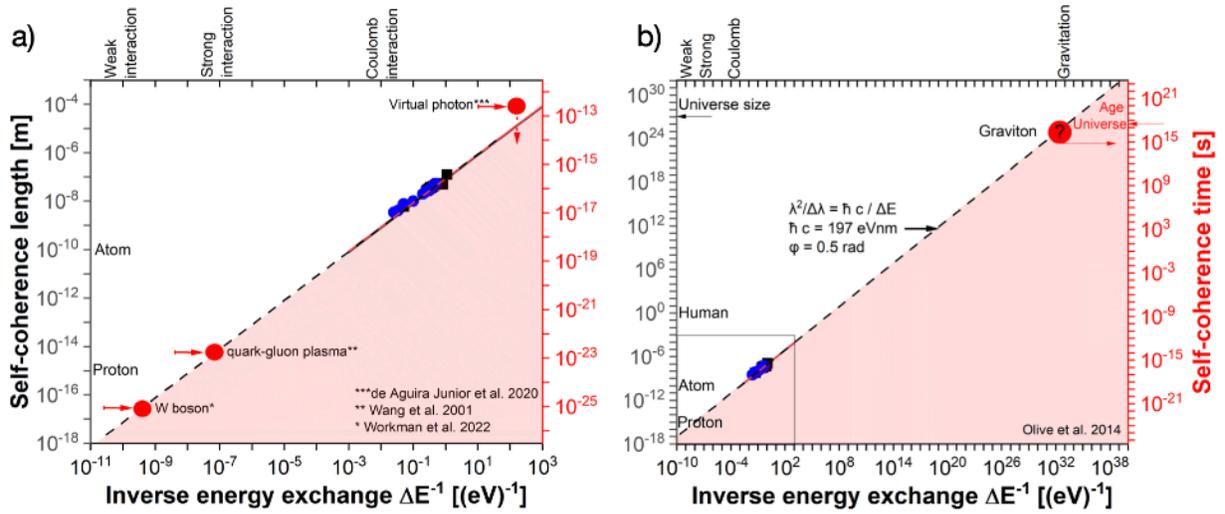

Figure 6: Comparison of lifetime measurements of virtual particles with self-coherence data. a) Weak, Strong and Coulomb interactions. b) Further scale expansion to include gravitational interaction. Red circles are lifetime data and blue symbols are self-coherence values of matter waves. Horizontal arrows capture the differences between energy and energy uncertainty. The vertical arrow is caused by instrumental limitations. The self-coherence scales are commented by selected physical objects.